\documentclass[]{agujournal2019}
\usepackage{url} 
\usepackage{amssymb}
\usepackage[inline]{trackchanges} 
\usepackage{soul}
\usepackage{xcolor}
\usepackage{graphicx} 
\usepackage{graphics}

\usepackage{multicol} 
\usepackage{amsfonts} 

\usepackage{mathpazo} 

\journalname{Geophysical Research Letters}

\begin{document}

\title{Dust impact voltage signatures on Parker Solar Probe: influence of spacecraft floating potential}

%
%



\authors{S. D. Bale\affil{1,2}, K. Goetz\affil{3}, J. W. Bonnell\affil{1}, A. W. Case\affil{4}, C. H. K. Chen\affil{5},T. {Dudok de Wit}\affil{6},
	L. C. Gasque\affil{1,2}, P. R. Harvey\affil{1}, J. C. Kasper \affil{7,4}, P. J. Kellogg\affil{3}, R. J. MacDowall\affil{8},
	M. Maksimovic\affil{9}, D. M. Malaspina\affil{10}, B. F. Page\affil{1,2}, M. Pulupa\affil{1}, M. L. Stevens\affil{4}, J. R. Szalay\affil{11}, A. Zaslavsky\affil{9}}

\affiliation{1}{Space Sciences Laboratory, University of California, Berkeley, CA 94720-7450, USA}
\affiliation{2}{Physics Department, University of California, Berkeley, CA 94720-7300, USA}
\affiliation{3}{School of Physics and Astronomy, University of Minnesota, Minneapolis, 55455, USA}
\affiliation{4}{Smithsonian Astrophysical Observatory, Cambridge, MA 02138 USA}
\affiliation{5}{School of Physics and Astronomy, Queen Mary University of London, London E1 4NS, UK}
\affiliation{6}{LPC2E, CNRS and University of Orl\'eans, Orl\'eans, France}
\affiliation{7}{Climate and Space Sciences and Engineering, University of Michigan, Ann Arbor, MI 48109, USA}
\affiliation{8}{Solar System Exploration Division, NASA/Goddard Space Flight Center, Greenbelt, MD, 20771}
\affiliation{9}{LESIA, Observatoire de Paris, Université PSL, CNRS, Sorbonne Université, Université de Paris, 5 place Jules Janssen, 92195 Meudon, France}
\affiliation{10}{Laboratory for Atmospheric and Space Physics, University of Colorado, Boulder, CO, 80303, USA}
\affiliation{11}{Department of Astrophysical Sciences, Princeton University, Princeton, NJ, 08544, USA}

\correspondingauthor{Stuart D. Bale}{bale@berkeley.edu}


\begin{keypoints}
\item The Parker Solar Probe (PSP) FIELDS instrument measures millisecond voltages impulses associated with dust impacts
\item The sign of the largest monopole voltage response is a function of the spacecraft floating potential
\item These measurements are consistent with models of dynamic charge balance following dust impacts
\end{keypoints}

Submitted :  \today
%
%

%
%


\begin{abstract}
When a fast dust particle 
hits a spacecraft, it generates a cloud of plasma some of which escapes into space and the momentary charge 
imbalance perturbs the spacecraft voltage with respect to the plasma.  Electrons race ahead of ions, however 
both respond to the DC electric field of the spacecraft.  If the spacecraft potential is positive with respect to the 
plasma, it should attract the dust cloud electrons and repel the ions, and vice versa.  
Here we use measurements of impulsive voltage signals from dust impacts on the Parker Solar Probe (PSP) spacecraft to 
show that the peak voltage amplitude is clearly related to the spacecraft floating potential, consistent with theoretical 
models and laboratory measurements.   In addition, we examine some timescales associated with the voltage waveforms 
and compare to the timescales of spacecraft charging physics.
\end{abstract}

\section*{Plain Language Summary}
When a fast, interplanetary dust particle hits a spacecraft, it generates a 
shock in the spacecraft material that liberates a hot, ionized plasma.  Some 
of the plasma ions and electrons return immediately to the spacecraft, but 
some escape into space.  The momentary charge imbalance created by the different 
ion and electron speeds generates a transient 
perturbation to the voltage of the spacecraft.  However, these electrons and ions can be attracted 
or repelled depending on the DC electric field of the spacecraft itself.  In this paper, 
we show this effect clearly:  when the spacecraft floating voltage is negative with 
respective to the interplanetary plasma, the electrons from the dust impact 
are repelled and vice-versa.

%
%

\section{Voltage signatures of dust impacts on spacecraft}

Plasma wave electric field measurements in space have proven to be powerful diagnostic of planetary \cite{1983JGR....88.8637S,1987JGR....9214959G,2009GeoRL..36.3103M,2014JGRA..119.6294Y} and 
interplanetary \cite{1997GeoRL..24.3125G,2012JGRA..117.5102Z,2014GeoRL..41..266M,2016JGRA..121..966K} dust processes.  A hypervelocity dust impact onto the spacecraft body or antenna produces a plasma cloud and the 
momentary charge imbalance generates a rapid perturbation to the spacecraft floating potential.  Some of the resulting plasma is recollected by the spacecraft and some of it escapes, depending on the energy of the ions and 
electrons and the spacecraft-to-plasma electric potential.  Monopole (probe-to-spacecraft) 
or dipole (probe-to-probe) voltage measurements will record millisecond-timescale spikes and/or their spectral content.  These effects have been explored in several recent papers \cite{2015JGRA..120..855Z,2017ITPS...45.2048V,2018JGRA..123.3273K,
2019JGRA..124.8179V,PJK2020} and recently reviewed by \citeA{2019AnGeo..37.1121M}.  

During the first PSP solar encounters, PSP/FIELDS measurements of dust impact rates were used to map the radial 
variation the flux of the interplanetary dust.  \citeA{2020ApJS..246...51P} used dipole voltage signals to infer the dust velocity vector and comparison 
with models \cite{2020ApJS..246...27S} suggests that this dust population is consistent with $\beta$ micrometeoroids on exiting hyperbolic 
orbits.  \citeA{2020ApJ...892..115M} compared data from Encounters 1-3 to show that the population is variable and probably has its 
source between 10-30 $R_S$.  While the PSP/WISPR instrument saw the beginnings of a decrease in F corona intensity \cite{2019Natur.576..232H}.

Here we examine the sign of the monopole antenna measurements from the PSP/FIELDS instrument; these observations show that 
the spacecraft voltage perturbation is influenced by the initial spacecraft potential 
itself.  If the spacecraft is initially negatively charged, it will attract more ions and repel 
more electrons from the plasma cloud producing a large positive perturbation.  If initially positively charged, 
the opposite occurs and returning electrons produce a large negative polarity spike.
We also examine some typical timescales associated with these perturbations and 
suggest that they are associated with the escape process.

\section{Parker Solar Probe Measurements}
The Parker Solar Probe (PSP) mission \cite{2016SSRv..204....7F} was launched in August 2018 into 
an orbit that will take it deep into the inner heliosphere with a final perihelion distance of 9.8 $R_S$ 
from the center of the Sun.  This study uses measurements from the PSP/FIELDS \cite{2016SSRv..204...49B} 
and the PSP/SWEAP \cite{2016SSRv..204..131K} instruments primarily from PSP Encounter 2, between March 23, 2019 
and April 13, 2020 to investigate the role of the spacecraft floating (DC) voltage on the voltage signature of 
dust impacts onto the spacecraft.  Perihelion of Encounter 2 was on April 5, 2019 at $\approx 35.7$ solar radii ($R_S$).

Our primary measurements are made by the Time Domain Sampler (TDS) subsystem of the PSP/FIELDS instrument \cite{2016SSRv..204...49B} .  
The TDS makes rapid samples of waveforms with simultaneous sampling of five analog channels which can be selected from dipole antennas pairs, 
monopoles, or a high-frequency search coil magnetometer \cite{2016SSRv..204...49B}.  During PSP Encounters 2 and 3 (used here), the TDS was 
configured to sample at 1.92 MSa/s and produce 32768-point waveform 'events'; therefore each TDS event is 17.067 ms in duration.  In addition to the 
TDS waveform events, the TDS records a 'TDS Max' value each 7 seconds during nominal encounter mode.  The TDS Max value is the signed extreme 
of the entire datastream during the interval and is dominated by the large voltage signatures of dust impacts.

We use TDS measurements of the voltage 
between the 'V2' monopole antenna and spacecraft ground $V_G$ and we call this measurement $\delta V_2 = V_2 - V_G$.  The $\delta$ emphasizes the fact 
that the TDS measurement is band-pass filtered (i.e. {\em not} DC-coupled); the TDS system has a flat gain and phase response from 1 kHz to 1 MHz (1 ms - 1 $\mu$s) and the 
waveforms shown here have not been corrected with a transfer function, which will not change our results on these timescales.  The 
measurement represents the voltage perturbation between the probe-spacecraft system.  Note that a positive perturbation of the spacecraft ground/potential 
at a fixed probe potential is measured as a negative voltage impulse in $\delta V_2$.  The V2 monopole is mounted near the plane of the spacecraft heatshield at $-\cos(55^\circ) \hat{x} - \sin(55^\circ) \hat{y}$ 
in the spacecraft coordinate system.  During PSP solar encounter, the spacecraft $\hat{x}$ axis points southward and the $\hat{y}$ axis points approximately in the ram direction, therefore the V2 monopole is on the anti-ram side during the normal solar encounter configuration \cite{2016SSRv..204...49B}. 
The PSP/FIELDS electric antennas are 2m long, 1/8" diameter thin-walled tubes of C103 Niobium alloy and have a 
free space capacitance of ~$C_A \sim$ 18pF and a system stray (base) capacitance of $C_B \sim$ 26pF \cite{2017JGRA..122.2836P}.
The (post-anneal) photoelectron threshold of C103 Niobium is $E_{103} \sim~$4.72 eV \cite{2019JSpRo..56..248D} and we assume a photoelectron temperature 
of $T_{ph} \sim 1.5$ eV.  

\begin{figure}[ht]
\noindent\includegraphics[width=\textwidth]{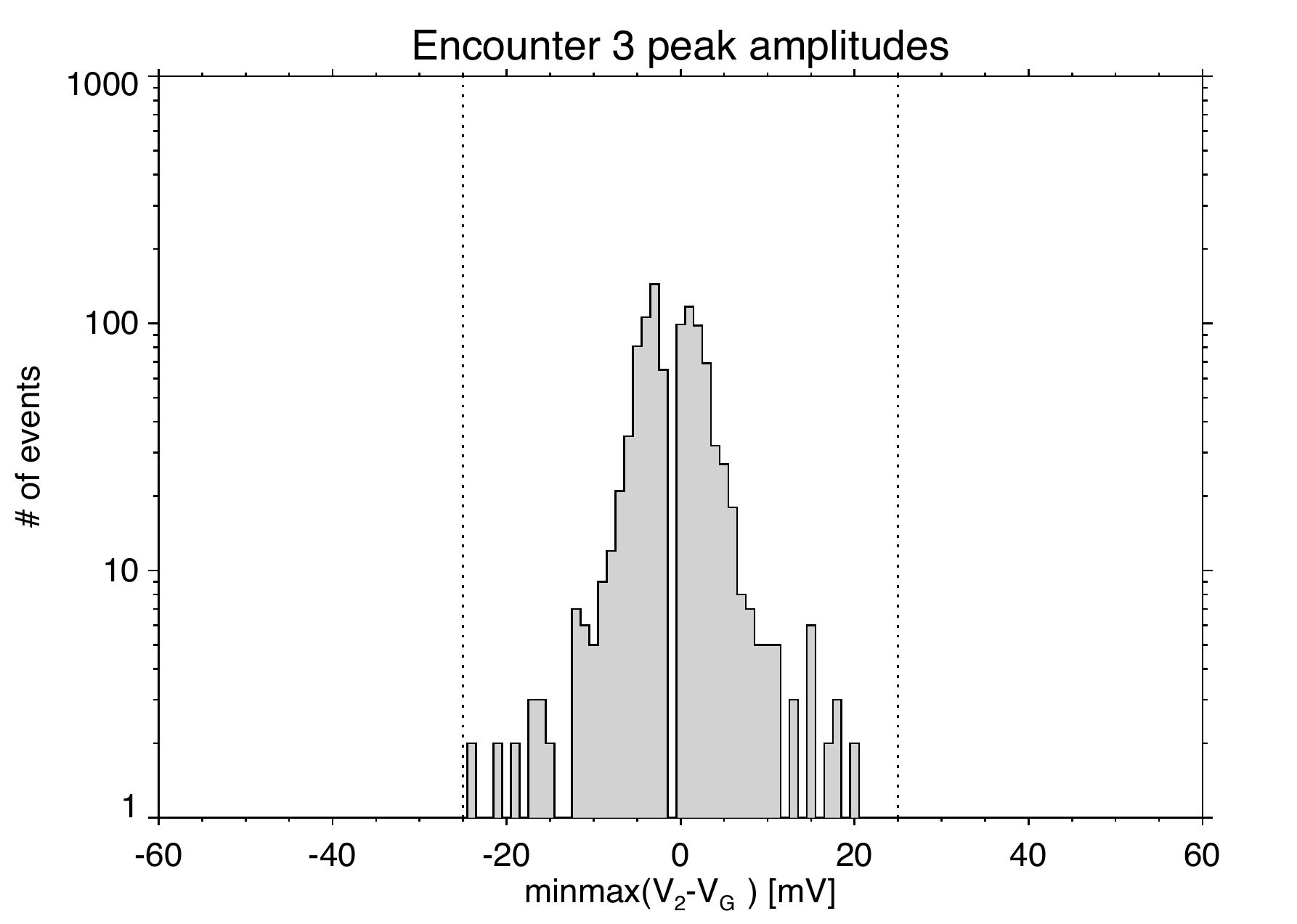}
\caption{Histogram of minimum and maximum waveform amplitudes on the $V2$ monopole during 
Encounter 3 on September 1, 2019 ($R$ $<$ 36 $R_S$), comprising 511 waveform events each represented 
here with a maximum and minimum data value.  Encounter 3 uses a burst waveform selection algorithm
that selects plasma waves, rather than dust impacts.  The measurements here represent a mix of ion acoustic, 
electrostatic whistler, and Langmuir waves.  Dotted vertical lines at $\pm$ 25 mV are our threshold 
for dust events during Encounter 2.}
\label{e3histo}
\end{figure}

For PSP Encounters 1 and 2, TDS was programmed to capture waveforms with the largest absolute amplitude and that algorithm returned predominantly dust impact events.  
For Encounter 3, a new algorithm was implemented that favors a combination of large amplitude {\em and} a large number of zero-crossings.  This algorithm generated primarily plasma wave 
events during Encounter 3.  Figure \ref{e3histo} is a histogram of both minimum and maximum amplitude from 511 waveform events on September 1, 2019 of Encounter 3 near 
perihelion.  All of these events are plasma waves, as identified by eye, and it can be seen that none exceed $\pm$ 25 mV amplitude.  We therefore choose $\pm$ 25 mV as our threshold beyond 
which we consider our measurements to be dust impacts, rather than plasma waves.  \citeA{2020ApJS..246...51P} used a threshold of 50 mV, which produces qualitatively similar results 
for this analysis, but fewer counts.

Our 'spacecraft potential' measurement $V_{SC}$ is computed as the negative of the average DC-coupled voltage on all four monopole antennas $V_{SC} = - (V_1 + V_2 + V_3 + V_4))/4$ 
from the Digital Fields Board (DFB) subsystem \cite{2016JGRA..121.5088M}.  The voltage probes $V_i$ are current-biased to hold them near the local plasma potential, therefore this quantity should represent the 
floating potential of the spacecraft \cite{1995AnGeo..13..118P,2012AnGeo..30.1075G}. However it is important to note that the PSP heatshield on the sunward side of the spacecraft is {\em not} directly electrically 
connected to the spacecraft body itself; electrical coupling between the heatshield and spacecraft body is carried by plasma and body currents only.  Therefore this measured $V_{SC}$ may not represent the 
true floating potential of the entire heatshield-spacecraft system.  Furthermore, modeling has suggested that for very high photoelectron densities, space-charge effects may produce a double layer 
at the sunward surface of the heatshield and a plasma wake behind that may modify its floating potential \cite{2010PhPl...17g2903E,2012AnGeo..30.1075G}.  Note that we differentiate $V_{SC}$ from $V_G$; $V_{SC}$ is the DC-coupled spacecraft 
potential measured by the DFB, while $V_G$ is the spacecraft potential reference for the TDS band-passed $\delta V_2$ measurement.

We also use electron (total plasma) density and electron core temperature measurements produced from an analysis of the quasi-thermal noise 
spectrum \cite{1989JGR....94.2405M} during PSP Encounter 2.  This technique was also applied to Encounter 1 data and described in \citeA{2020ApJS..246...44M}.  
Ion temperature measurements are derived from SWEAP instrument moment calculations of the total ion distribution from the Solar Probe Cup instrument \cite{2016SSRv..204..131K,2020ApJS..246...43C}.

\section{Results}

\begin{figure}[ht]
\hspace*{-2cm}\includegraphics[width=18cm]{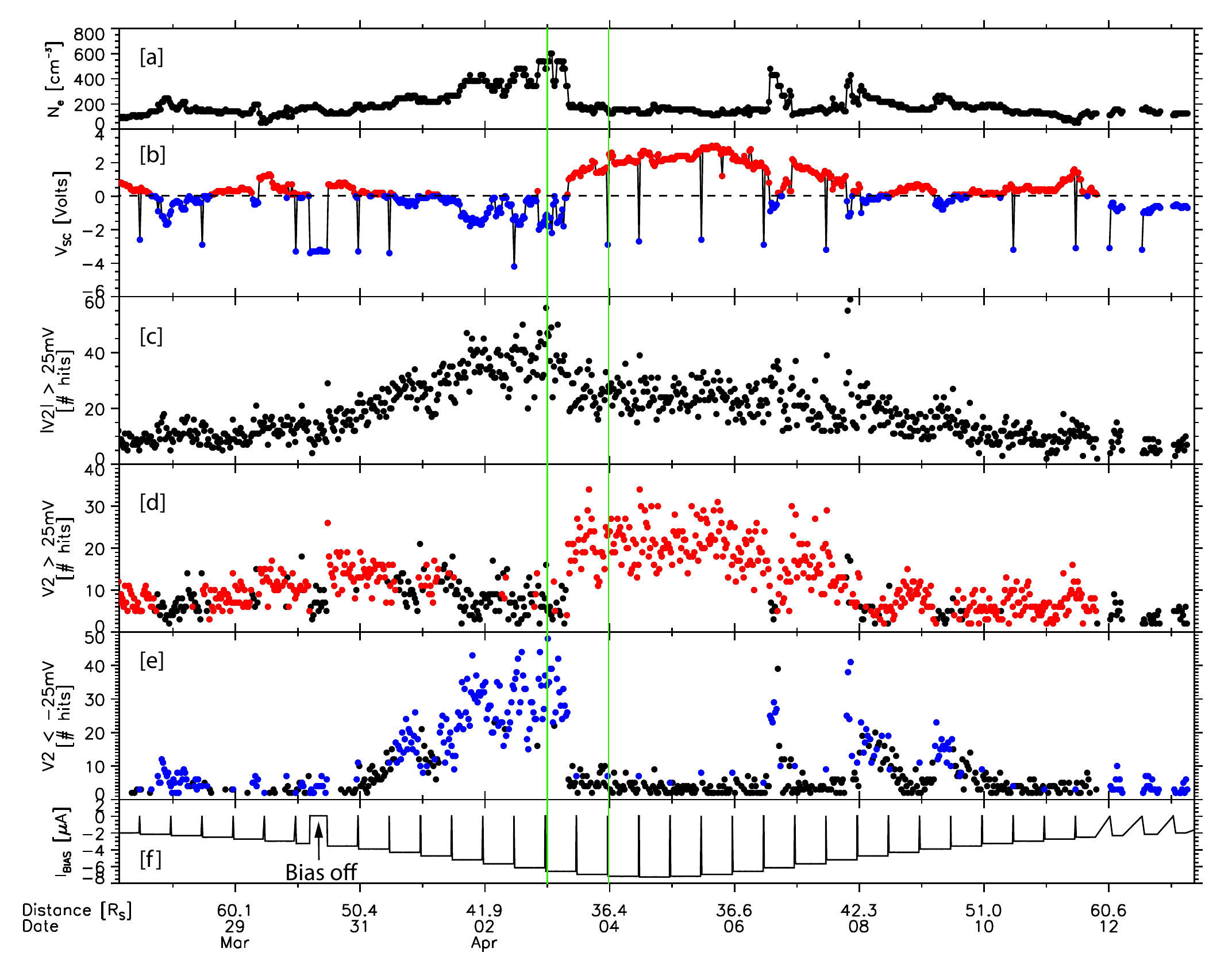}
\caption{An overview of PSP Encounter 2 data, with perihelion at 35.7 $R_S$ on April 5, 2019.  Panel [a] 
is the total plasma density computed from the QTN spectrum.  Panel [b] is the spacecraft potential proxy $V_{SC}$, 
colored for polarity (red $>$ 0, blue $<$ 0).  Panel [c] is the number of dust hits in 30 minute intervals estimated from $|V_2|$, 
showing an increase towards perihelion.  Panel [d] is the number of dust hits in 30 minute intervals with positive polarity, where 
red dots are intervals of $V_{SC}$ $>$ 0.  Panel [e] is the number of dust hits in 30 minute intervals with negative polarity, where 
blue dots are intervals of $V_{SC}$ $\leq$ 0.  Panel [f] is the value of the applied bias current in $\mu A$.  Spikey events are twice-daily 
bias sweeps, and an interval of no bias is indicated.  Green vertical bars show the interval of April 3, 2019 with a large change of 
density and spacecraft potential/polarity.  This figure shows clearly that intervals of negative (positive) $V_{SC}$ correspond to clear enhancements in 
dust impulses with positive (negative) dominated TDS events; note that TDS measures $V_2$ - $V_{G}$, so that rapid voltage changes in $V_{G}$ appear 
opposite polarity.}
\label{e2summary}
\end{figure}

Figure \ref{e2summary} shows an overview of PSP Encounter 2 measurements from March 27 to April 13, 2019.  The top panel [a] is the total plasma density 
$n_e$ measured from the plasma frequency peaks in the quasi-thermal noise spectrum \cite{1989JGR....94.2405M,2020ApJS..246...44M}.  The density reaches 
a peak value of around $n_e \sim$ 600 cm$^{-3}$ on April 3 at r $\sim$ 39 $R_S$.  Panel [b] is the spacecraft potential $V_{SC}$ as described in Section 2 above, 
colored for polarity (red $>$ 0, blue $<$ 0), the values  range between -4 and +4 Volts.  Note that higher plasma density intervals correspond to more negative spacecraft 
floating potential, as expected \cite{1995AnGeo..13..118P}. Panel [c] shows the number of dust events in 30 minute intervals, assuming that TDS Max values of $|V_2| >$ 25 mV are attributable to a dust impact and shows the trend of increasing dust impacts with radial distance as shown previously \cite{2020ApJS..246...51P,2020ApJ...892..115M,2020ApJS..246...27S}.  Panel [d] is the number of positive polarity dust events ($V_2 >$ 25 mV) per 30 minutes, with red dots indicating intervals of positive $V_{SC}$ from panel [b] and panel [e] is the number of negative polarity dust events ($V_2 <$ -25 mV) with blue dots indicating intervals of negative spacecraft potential.
The bottom panel [f] shows the bias current applied to the V2 antenna, indicating an interval of no bias, as well as
times when the fixed bias was disabled during bias current calibration sweeps.  The
sweep intervals were deleted from the statistics to keep the sweeps from
contaminating the dust impact measurements.

 The striking feature of Figure \ref{e2summary} is the relationship  
between the sign of the spacecraft potential in panel [b] and the sign of the dust voltage impact signal in panels [d] and [e]; negative polarity spacecraft potential tends to 
produce negative impulse TDS events $\delta V_2 = V_2 - V_{G}$, i.e. positive perturbations to $V_G$ assuming a fixed $V_2$.  This is consistent with the idea that a negatively  
charged spacecraft $V_{SC} <0$ repels the dust cloud electrons producing a positive voltage perturbation $\delta V_G$ (therefore a negative perturbation to $\delta V_2 = V_2 - V_G$), 
and vice-versa.  Note that this behavior is consistent with plasma cloud ions and electrons having temperatures on the order of $\sim$1 V, as described further below.

Note also that in some cases of high density (panel [a]) and therefore negative $V_{SC}$ (panel [b]), there appears to be enhanced overall count levels (panel [c]), suggesting 
that the spacecraft floating voltage may influence or modify the overall dust impact rate by perhaps providing a threshold for the measured charge.

\begin{figure}[ht]
\noindent\includegraphics[width=\textwidth]{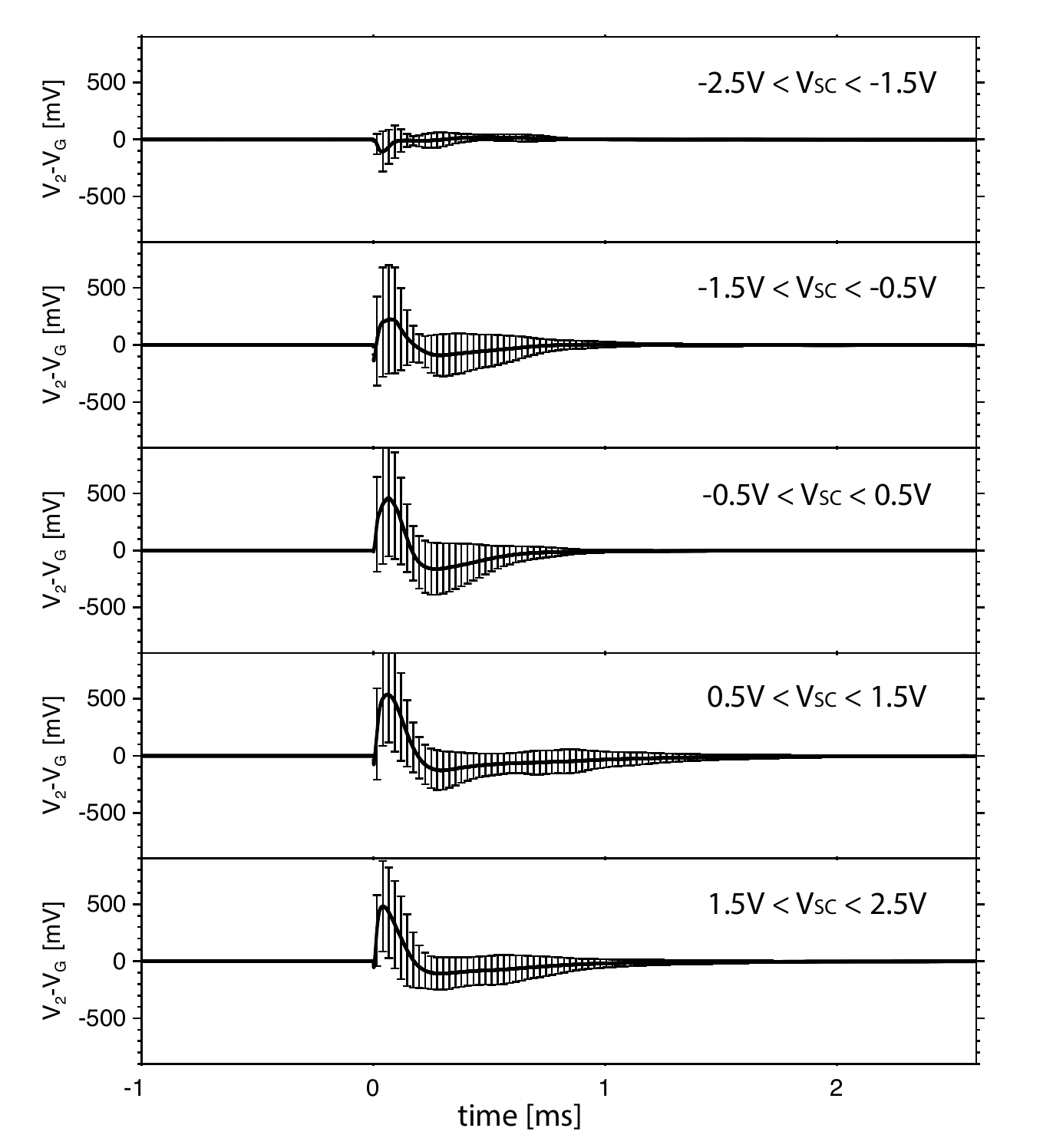}
\caption{Superposed epoch plots of V2 monopole TDS waveform events, organized by spacecraft floating voltage $V_{SC}$.  
The black curve is the average waveform and error bars are one-sigma variations.  The instrument saturates at $\approx$ $\pm$ 1100 mV.  Table \ref{tdstable} collects the parameters associated 
with these intervals.  Notably, the top panel with $V_{SC}$ $<$ -1.5V shows that the typical waveform has its peak value with $V_2$-$V_{SC}$ $<$ 0.}
\label{tdsstack}
\end{figure}

\begin{figure}[ht]
\noindent\includegraphics[width=\textwidth]{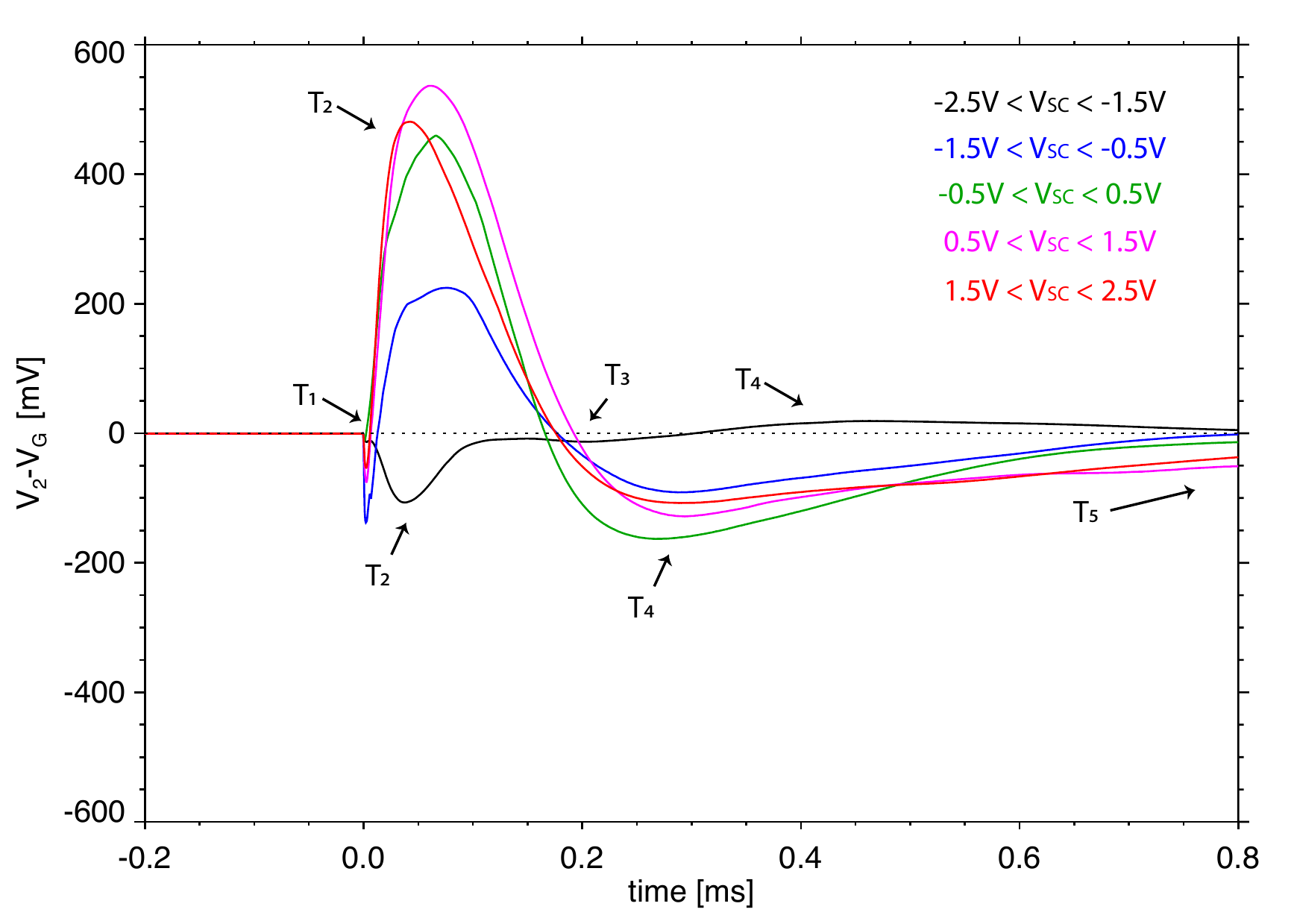}
\caption{Epoch-averaged waveforms binned and colored by spacecraft potential.  These are the average waveforms from Figure \ref{tdsstack}.
Note that the waveform with $V_{SC}$ $<$ -1.5V is opposite polarity to the others at both the primary peak at $T_2$ and the secondary peak at $T_4$.  Table \ref{tdstable} 
collects the waveform and plasma parameters.}
\label{tdsall5}
\end{figure}

This effect can also be seen by examining the waveforms themselves; we use waveforms from April 3, 2019, the interval between green vertical 
bars in Figure \ref{e2summary}.  Figure \ref{tdsstack} shows superposed-epoch averaged 
waveforms, with 1$\sigma$ deviations, and organized by intervals of spacecraft potential $V_{SC}$.  The number of waveforms used to compute 
the average and variance in each row is given in column 2 of Table \ref{tdstable}.  Each individual waveform is shifted to start at t=0 
by finding the initial perturbation above $5 * V_{noise}$, where $V_{noise}$ is the RMS level of the first 4 ms of the waveform (before the 
dust signal).  This level is typically $V_{noise} \approx$ 0.5 mV.  While there is considerable variability 
between the waveforms, they generally exhibit some common features:  an initial rapid negative excursion, a peak at $T \lesssim$ 0.1 ms, and 
another peak at 0.2 $\lesssim T \lesssim$ 0.5 ms.  The voltage impulse settles back to $V_{noise}$ at times $T \gtrsim$ 2 ms.  
The average waveforms are plotted together in Figure \ref{tdsall5} and colored for range of $V_{SC}$, with several timescales annotated and 
collected in Table \ref{tdstable}.

\begin{table*}[ht]
\small
\caption{Parameters associated with TDS epoch-averaged waveforms in Figure ~\ref{tdsall5} and plasma parameters as described in the text.}
 \centering
 \resizebox{\columnwidth}{!}{
  \begin{tabular}{c|c |c| c|c|c|c|c|c|c|c|c|c}
 \hline
  $ V_{SC}$  & \#  & $\langle n_e \rangle$   & $\langle T_e\rangle$& $\langle T_i\rangle$ & $\langle\tau_{pe}\rangle$ &$\langle\tau_{ce}\rangle$ & $\langle\tau_{SC}\rangle$&$T_1$ [ms]&$T_2$ [ms]&$T_3$ [ms]&$T_4$ [ms]&$T_5$ [ms]\\
   \  $[V]$ &  &$[cm^{-3}]$ &  $[eV]$&  $[eV]$ &[ms]&[ms]&[ms]&1st z/c&2nd peak&2nd z/c&3rd peak&$\rightarrow V_{noise}$\\
 \hline
   $(-2.5, -1.5)$  & 19  & 571 &25  & 9 &0.005&0.436&0.244& - & 0.037&0.310&0.463&3.346\\
   $(-1.5, -0.5)$  & 31  & 515 & 25 & 10 &0.005&0.448&0.256&0.013& 0.076&0.176&0.286&4.208\\
   $(-0.5, 0.5)$  & 19  & 352 & 32 & 13 &0.006&0.403&0.421&0.002&0.066&0.165&0.266&3.532\\
   $(0.5, 1.5)$  & 57  & 177 & 28 & 11 &0.008&0.369&0.797& 0.006& 0.061&0.192&0.294&2.786\\
   $(1.5, 2.5)$  & 84  & 163 & 30 & 11 &0.009&0.366&0.927&0.006&0.043&0.175&0.288&2.395\\
 \hline
 \end{tabular}
 }
\label{tdstable}
 \end{table*}
 
The columns in Table \ref{tdstable} list 
 (left-to-right) the interval of $V_{SC}$ in Volts, the number of waveform events in that interval, the average plasma density $\langle n_e \rangle$ , electron core temperature $\langle T_e \rangle$, 
 ion temperature $\langle T_i \rangle$, the electron plasma period $\langle \tau_{pe}\rangle$, electron cyclotron period $\langle \tau_{ce} \rangle$, the spacecraft RC time $\langle \tau_{SC}\rangle$ and timescales associated with the 1st zero-crossing, 
 2nd peak, 2nd zero-crossing, 3rd peak, and relaxation to $V_{noise}$ (zero) respectively, as annotated in Figure \ref{tdsall5}.  In addition, all waveforms with $V_{SC} > -1.5V$ show a small (negative) 1st peak at $T \approx T_1/2$, although these values start to approach the granularity of the measurement $1/$(1.92 MSa/s) $\sim ~0.53 ~\mu$s. 
The antenna RC charging time can be estimated as $\tau_{A} \simeq {(C_A T_{ph})}/{I_{bias}}$, which takes values of 0.01 to 0.04 ms during Encounter 2 and is generally 
less than the spacecraft RC charging time $\tau_{SC} \simeq (C_{SC} T_{e})/(A_{SC} ~n_e e ~v_{th,e} )$ column 8 in Table \ref{tdstable}.  We estimate $C_{SC}$ as the free space capacitance of a 1m radius sphere $C_{SC} \approx$ 110 pF.  Note that we use the thermal electron 
temperature to estimate $\tau_{SC}$, rather than the photoelectron temperature \cite{2015JGRA..120..855Z} since the spacecraft body is not illuminated; using $T_{ph}$ would result in smaller values of $\tau_{SC}$ by a factor of $\sim$10.

Models of dust waveform signatures associate characteristic timescales with the ion and electron collection and escape processes \cite{2015JGRA..120..855Z,2017JGRA..122....8M,
PJK2020,Shen2020}.  In particular, escape timescales $\tau_e \sim R_{SC}/v_e$ and $\tau_i  \sim R_{SC}/v_i$ can be estimated \cite{Shen2020} using impact cloud electron and ion temperatures $v_e$ and $v_i$ \cite{2016JGRA..121.8182C}, where $R_{SC} \sim$ 1m is the spacecraft scale size.  If we assign the values of our first peak $T_1/2$
with an electron timescale $R_{SC}/v_e$ and  our second peak $T_2$ with an ion timescale $R_{SC}/v_i$ we find escape speeds of $v_e \approx$ 150-1000 km/s ($\sim$0.3  - 3 eV) and $v_i \approx$ 15-30  km/s ($\sim$1-5 eV, assuming protons).  These energies are consistent with laboratory measurements of dust cloud plasma temperatures \cite{2016JGRA..121.8182C}.  Again, note that there is little or no peak at $T_1/2$ for the waveform with $V_{SC} < -1.5$V (the most negative values).
Note that the timescales associated with the dust impacts arise from charging interactions between the electrons, ions, and the spacecraft, which will be convolved to produce the probe-spacecraft voltage impulse.  Our analysis here 
is very simple and only to suggest that there are multiple timescales evident in the measurement and they can be associated with electron and ion motion.  The voltage amplitudes of the waveforms are also interesting and presumably 
related to the charge generated during impact and the fractions collected and repelled.

\section{Conclusions}

We show that the sign of the voltage waveforms and onboard voltage extrema are influenced by the DC spacecraft floating potential 
in a way that is consistent with models of dust plasma cloud dynamics.  Timescales associated with the waveforms are broadly consistent 
with expectations for electron and ion escape energies, and those in turn are at appropriate energies to be influenced by the $\mathcal{O}(\pm$ 2V) 
spacecraft potential measurement.  As noted above, the measured $V_{SC}$ may not represent the true floating potential of entire spacecraft-heatshield system; as our models 
of dust plasma cloud escape become more sophisticated, these waveforms may add insight to the actual floating potential of the PSP 
spacecraft-heatshield system.

More detailed, and statistical, analysis of the PSP/FIELDS dust data is likely to add insight to the physics of dust plasma cloud dynamics, and well as the 
global distribution of interplanetary dust.


%

%

\acknowledgments
The FIELDS experiment on the Parker Solar Probe spacecraft was designed and developed under NASA contract NNN06AA01C. The PSP/FIELDS team acknowledges the extraordinary contributions of the Parker Solar Probe mission operations and spacecraft engineering teams at the Johns Hopkins University Applied Physics Laboratory.  SDB acknowledges the support of the Leverhulme Trust Visiting Professorship program. Data access and processing was done using the SPEDAS IDL environment \cite{2019SSRv..215....9A}.  PSP/FIELDS data is publicly available at \url{http://fields.ssl.berkeley.edu/data/}.


\newpage
\bibliography{./dust}

%
%
%
%
%

\end{document}